\documentclass[acmsmall,screen,nonacm]{acmart}

\AtBeginDocument{%
  }

\usepackage{xspace}
\usepackage{listings}
\usepackage{paralist}
\usepackage{wrapfig}
\usepackage{xurl}
\usepackage{hyperref}

\newcommand\maybecolor[1]{\color{#1}}

\let\ls\lstinline
\let\li\lstinline

\newif\iffull\fullfalse    
\newif\ifdraft\drafttrue

\newif\ifanon\anontrue 

\ifdraft
\newcommand\comm[3]{{\color{#1}[\emph{#2 says:} {#3}]}}
\newcommand\todo[1]{{\color{dkred}[\emph{TODO: }{#1}]}}
\else
\newcommand\comm[3]{}
\newcommand\todo[1]{}
\fi

\definecolor{addition}{rgb}{0,0.1,0.5}
\definecolor{dkblue}{rgb}{0,0.1,0.5}
\definecolor{dkgreen}{rgb}{0,0.4,0}
\definecolor{dkred}{rgb}{0.6,0,0}
\definecolor{dkpurple}{rgb}{0.7,0,1.0}
\definecolor{purple}{rgb}{0.9,0,1.0}
\definecolor{olive}{rgb}{0.4, 0.4, 0.0}
\definecolor{teal}{rgb}{0.0,0.4,0.4}
\definecolor{azure}{rgb}{0.0, 0.5, 1.0}
\definecolor{gray}{rgb}{0.5, 0.5, 0.5}
\definecolor{dkgrey}{rgb}{0.2, 0.2, 0.2}
\definecolor{lilac}{rgb}{0.70, 0.04, 0.08}
\definecolor{applegreen}{rgb}{0.55, 0.71, 0.0}

\newcommand{\secref}[1]{\mbox{Section~\ref{#1}}\xspace}

\newcommand{\kw}[1]{\ensuremath{\mathsf{#1}}}

\newcommand\fstar{F$^\star$\xspace}
\newcommand\krml{KaRaMeL\xspace}
\newcommand\name{StarMalloc\xspace}
\newcommand\hm{hardened\_malloc\xspace}

\newcommand\hoare[3]{\ensuremath{\{~#1~\}~#2~\{~#3~\}}}

\newcommand\ptsto[3][]{\ensuremath{\kw{#2} \stackrel{\kw{#1}}{\mapsto} \kw{#3}}}
\newcommand\estar[2]{\ensuremath{#1~\ast\,~#2}}
\newcommand\sref[1]{Section~\ref{sec:#1}\xspace}
\newcommand\fref[1]{Figure~\ref{fig:#1}\xspace}
\newcommand\tref[1]{Table~\ref{tab:#1}\xspace}
\newcommand{\myparagraph}[1]{\vspace{0.5em}\noindent\textbf{#1.}~}

\begin{document}

\title[\name: A Formally Verified, Security-Oriented Memory Allocator]{\name: A Formally Verified,
  Concurrent, Performant, and Security-Oriented Memory Allocator}

\author{Antonin Reitz}
\affiliation{%
  \institution{Inria}
  \city{Paris}
  \country{France}
}
\email{antonin.reitz@inria.fr}

\author{Aymeric Fromherz}
\affiliation{%
  \institution{Inria}
  \city{Paris}
  \country{France}
}
\email{aymeric.fromherz@inria.fr}

\author{Jonathan Protzenko}
\affiliation{%
  \institution{Microsoft Research}
  \city{Redmond}
  \country{USA}
}
\email{protz@microsoft.com}

\renewcommand{\shortauthors}{Reitz et al.}

\begin{abstract}
Despite the rise of memory-safe languages such as Rust or Go, most safety-critical,
systems-oriented, low-level applications are still written in unsafe languages like C and C++. The
reason is simple: the boundary between the OS and userland is fundamentally thought of, designed and
explained in C terms through the various headers and man pages that capture the calling conventions,
and thus, enshrine the C ABI as the correct level of abstraction when writing systems-oriented
code.
The system memory allocator (\li+malloc+, \li+free+\dots) is a prime example. Sitting at the boundary
between the OS and userland, the behavior of the system allocator was set in stone in 1989~in the C
standard, and has fundamentally not changed since then. And due to its very low-level nature, an
allocator must directly interface with the OS and juggle various low-level system APIs that are best
accessed from C -- indeed, such low-level memory management would hardly pass Rust's borrow checker!

But of all the low-level, safety-critical systems-oriented programs, the allocator is one of the
very few hit with a proverbial \emph{double whammy}. First, the allocator must assume the inevitable
reality of software development: most client programs will be written in unsafe languages and as a
result will be excruciatingly buggy. This means the allocator must go great lengths to include
security mitigations against common heap vulnerabilities. But as a result a second penalty arises:
the allocator \emph{itself} is not only increasingly complex, but must remain written in exactly the
same kind of memory-unsafe languages (e.g., C) to enjoy access to the OS APIs that it needs (e.g.,
\li+mmap+). Unsurprisingly, even widely-used and well-audited
allocators, such as the one in glibc, have been the victim of bug reports in recent years.
Given their security-critical aspect, allocators
thus require strong, formal guarantees about their correctness and safety.

In this work, we present \name, a verified, security-oriented, concurrent
memory allocator that can be used as a drop-in replacement in real-world
projects. Using the Steel separation logic framework, we show how to
specify and verify \name, relying on dependent types and modular
abstractions to enable efficient verification. As part of \name, we also
develop several generic datastructures and proof libraries directly reusable
in future systems verification projects. We finally show that \name
can be used with real-world projects, including the Firefox browser, and evaluate it
against 10 state-of-the-art memory allocators, demonstrating its competitiveness.
\end{abstract}

\maketitle

\section{Introduction}

In recent years, memory-safe languages such as Rust and Go have significantly
risen in popularity and adoption, and to this date remain some of the
fastest-growing languages~\cite{github-report}. However, for safety-critical,
low-level applications,
C and C++ remain the languages of choice, owing to their
ability to interact with the operating system and the memory in extremely
low-level ways.
While C and C++ allow developers to achieve high performance,
particularly in resource-contrained environments, C and C++ are also fundamentally
memory-unsafe, meaning they empower developers with the ability to ``shoot
themselves in the foot'' with great ease, resulting in a rich cornocupia of
well-known attack vectors, and especially of memory corruption bugs~\cite{sokmemory}.
Indeed, recent studies by Google~\cite{chromiummemorysafety} and Microsoft~\cite{microsoftmemorysafety}
estimate that memory-related errors, e.g., buffer overflows or dangling pointers,
are the root cause of 70\% of the security vulnerabilities in
widely deployed software.

Since no one in their right mind would expect the software industry to rewrite
everything (and especially legacy code) in Rust, we now ask: what can be done to mitigate the impact of
incorrect memory management in existing software codebases?
As argued by \citet{berger2012seatbelts}, ``the software industry is in a position similar to that
of the automobile industry of the 1950s, delivering software with lots of horsepower and tailfins
but no safety measures of any kind''.
One particular such airbag is the memory allocator, which oftentimes is the
last line of defense against incorrect memory management performed by the
client program. Realizing this, modern memory allocators~\cite{novark2010dieharder,
berger2006diehard,leijen2019mimalloc,hardenedmalloc,silvestro2017freeguard,scudoallocator}
often include mitigations against common heap vulnerabilties, for instance by randomizing allocations,
or separating heap metadata from the heap itself. The net effect is that attacks
are harder to conduct, and as such the allocator successfully provides some degree
of protection against incorrect programs.

Alas, allocators themselves are not immune to the kind of bugs they are supposed
to defend against! They are typically written in C/C++; and because they stand
on the critical path of most, if not all, client programs,
allocators also must fulfill numerous goals~\cite{wilson1995allocsurvey}, such
as: high performance; low memory consumption; maximizing concurrency; minimizing
heap contention; and so on. Meeting these goals requires custom data structures
and low-level pointer and bit manipulations, and typically
entails bugs: even widely used and audited allocators such as glibc are not immune, and several issues
were reported in recent years~\cite{glibcbug2020,hardenedmallocundefbug2021,glibc26306,
glibc22343,glibc22375,glibccve20134332,sqlbug94647,badalloc}.

Because allocators are so critical, they deserve, in our opinion, the highest
degree of assurance; that is, allocators ought to be formally verified, in order
to guarantee that no matter the input, they will always correctly and safely
maintain their data structures and invariants, hence functional correctness.
Formal verification has been successfully applied to a variety of application
domains, such as compilers~\cite{compcert,compcert-ct,cakeml}, operating
systems~\cite{gu2016certikos,sel4-sosp,klein2018formally}, and
cryptography~\cite{evercrypt,haclxn,fiat-crypto,jasmin20,appel15sha256,hmac-drbg}.
Formal verification has also been applied to memory allocators~\cite{sahebolamri2018allocator,mangano2017formal,jiang2019formally,
marti06osverif,refinedc,tuch07seplogic,gu2016certikos,appel2020verified}.
But to the best of our knowledge, formal verification was never applied to a
real-world allocator, complete with advanced book-keeping data structures,
sharded allocation pools and performance optimizations, fine-grained
concurrency, defensive security mitigations, and so on; in short, there is no
verified, state-of-the-art modern memory allocator.

We posit there are several reasons for this glaring omission from the formal verification landscape.
First, the secure allocator community has little overlap with the PL community, meaning there are
many well-established, peer-validated allocator designs, but they have yet to be subjected to formal
verification; in other words, the difficulty today lies not in proving a design secure, but rather
in proving its \emph{implementation} secure. As a consequence, a second possible explanation is that
verifying a real-world, \emph{system} allocator is a veritable challenge. The allocator sits below
the usual level of abstraction of typical verification frameworks; requires extensive modeling of
the environment, and notably the behavior of a variety of OSes and their APIs that each vary in
slightly non-POSIX ways, a notoriously tedious task; and requires verifying data structures with
punishing amounts of complexity, entangling as many as five linked lists together. For all of these
reasons, we are not aware of any \emph{real, secure, verified} allocator that is simultaneously performant,
features the latest security features, \emph{and} can act as a drop-in replacement for \li+malloc+,
\li+free+, and the myriad of slightly non-standard auxiliary APIs that must be implemented
nonetheless for ``real-world'' use.

In this work, we propose \name, the first verified, real-world, efficient,
security-oriented, concurrent memory allocator. Our design is based on
\hm~\cite{hardenedmalloc}, a scalable, security-focused, general purpose memory
allocator with a focus on long-term performance and memory usage. Doing so, we
leverage the latest advances in allocator design, embrace a solution that has met the approval of
the security community and whose merits are not up for debate, all the while providing a state-of-the-art
\emph{and} verified allocator. \name comes as C code; our theorems show that we
not only avoid memory errors (such as out of bounds accesses within the
allocator's internal data structures), but also functionally behave like an
allocator, even with a complex design (slabs, separate metadata, contiguous size
classes, bitmaps and bit-level arithmetic, doubly-linked lists within a static
array, AVL trees with constant-time metadata access, FIFO queues) that
incorporates additional security mechanisms (zeroing, guard pages, quarantine).

\name is authored and verified using the \fstar/Steel~\cite{mumon,steel2021} separation logic
framework, and is readily available as a C library. Our code is open-source, and
extensive testing has shown that \name can provide a drop-in replacement for
the system allocator on real-world programs.
Our contributions are the following:
\setdefaultleftmargin{0em}{}{}{}{}{}.
\begin{compactitem}
\item A new specification logic specification of a memory allocator and its security defense
  mechanisms
\item A collection of base libraries for the verification of
  low-level, raw memory-manipulating programs
\item A modular approach to verification with multiple layers that leverages
  dependent types and ML-style abstractions
\item A verified implementation of a security-oriented allocator, \name
\item An evaluation against 10 existing memory allocators on widely used
  benchmarks and real-world projects, including Mozilla Firefox, showing performance to be competitive.
\end{compactitem}

\section{Related Work}

\subsection{Toward Secure Memory Allocators}

Dynamic memory allocation in its current form stems from the original 1989
version of the C standard which specifies \li+malloc+, \li+calloc+,
\li+realloc+ and \li+free+. Since 1989, the purpose has remained the same: to
provide a \emph{portable}, abstract interface between the OS and the client
program. Each operating system exposes a system call that allows a process to
request more memory (originally \li+brk+, later on, with the advent of virtual
memory, \li+mmap+ or \li+VirtualAlloc+). Relying on this OS-specific system
call, the allocator grows the heap as needed (to avoid too-frequent syscalls)
and maintains internal book-keeping data structures that allow it to service the
allocation and de-allocation requests from the client program. The client
program, by virtue of relying on the system-provided dynamic memory allocator,
is happily unaware of all of these implementation details.

Even though the mandate of an allocator is simple, the design space for
possible implementation is large~\cite{wilson1995allocsurvey}. As noted in
the introduction, the allocator must satisfy several constraints that are
oftentimes at odds with each other: low fragmentation, good multithreaded
behavior, cache locality, great performance, and of course, security. What
constitutes a good design also changes over time: one notable inflexion
point in the design of allocators happened in the late 2000s, when better
benchmarking increased awareness of
fragmentation; combined with the generalization of multi-processor systems, this
led to a new generation of allocators, the most notable of which is
\li+jemalloc+~\cite{evans2006jemalloc}, which became the standard in Mozilla Firefox and
many BSDs.

More recently, the seemingly never-ending stream of memory
vulnerabilities led to new designs: state-of-the art allocators not only
enjoy high performance, but also now incorporate security at their core.
Security concerns led to typed memory allocators~\cite{
vanderkouwe2018typedalloc,akritidis2010cling}, allocators
for tagged memory (e.g., CHERI~\cite{bramley2023picking,watson2019cheri}),
or hardened memory allocators~\cite{hardenedmalloc,novark2010dieharder,
berger2006diehard,leijen2019mimalloc,silvestro2017freeguard,scudoallocator,muslmalloc}.
In short, designing a ``good'' memory
allocator remains an active research area, and incorporating defensive security
measures is now \emph{de rigueur}.

\subsection{Verified Memory Allocators}
Given their critical aspect, userspace memory allocators have long been considered
an interesting target for formal verification.
Using the Isabelle/HOL proof assistant~\cite{nipkow2002isabelle},
\citet{sahebolamri2018allocator} verify a simple sequential memory allocator
where a linked list intertwined with data
keeps track of allocation units. This is akin to the original design explained in
the K\&R book~\cite{kandr}, but a far cry from what modern allocators do.
\citet{appel2020verified} verify an array-of-bins malloc/free
system using the Verified Software Toolchain~\cite{vst}
embedded in Coq. Their implementation does not
support concurrency (it is single-threaded), and is intended
for use with verified clients, meaning it is unsuitable in adversarial settings.
They do, however, add extra reasoning in Coq to model resource awareness and
exhaustion.
\citet{refinedc} develop a methodology to verify C code using the Iris
framework~\cite{iris15,iris17,jung2018iris}, and apply it to the verification
of a small thread-safe allocator consisting of 68 lines of code.
An illustrative example fits within 68 lines of code, but a modern memory
allocator that meets state-of-the-art, ``real-world'' constraints needs more
than that.
\citet{wickerson2010explicit} use
rely-guarantee~\cite{jones1983tentative}, a verification
technology that predates the widespread adoption of separation logic, to verify
the Unix Version 7 memory manager. Unix version 7 was released in 1979, and its
memory manager relies on \li+sbrk+, which according to its \li+man+ page is a
``historical curiosity''.
Many other works~\cite{fang2018formal,yu2003building} verify \emph{simplistic}
memory allocators, but to the best of our knowledge, none verify a
state-of-the-art, competitive, real-world allocator. One exception
is~\citet{zhang2019verified}, which verifies a non-trivial allocator design,
except they operate on a \emph{specification}, and have no actual
implementation.

Perhaps more importantly, none of these allocators are intended to act as \emph{drop-in}
replacements for the system allocator, and have not been evaluated as such. Based on a combination
of careful examination of the papers above, and personal communications with some of the authors, it
is our understanding that none of these allocators ever intended to be usable in the real world.
This is by no means shocking: providing a system allocator requires implementing a wide API surface
(both POSIX \emph{and} non-standard but widely used APIs), while also deploying sophisticated
implementation techniques to provide performance and security. In short, all of the userspace
allocators we are currently aware of are proof-of-concepts, written without performance or security
in mind, and as such without any ambition to act as deployable system components.

Beyond userspace allocators, previous verification projects aimed to provide
verified operating systems and microkernels, which thus required kernel space
memory management~\cite{marti06osverif,tuch07seplogic,gu2016certikos,mangano2017formal}.
Being located in the kernel, these allocators have different constraints. First,
kernels have a lot of static data structures so these allocators do not need to
service as many small heap-based allocations; second, kernels oftentimes have
global locks for data structures~\cite{peters2015microkernel}, meaning concurrency
isn't as much of a concern for the allocator.
Furthermore, in the context of verified OS kernels like above, additional proofs
ensure that the in-kernel allocator does not need to be defensive against other
parts of the kernel, which means even in this context, a full-fledged defensive
general-purpose security-oriented allocator was not typically the object being studied.

One last area of related work to note is that of automatic memory management
runtime systems, i.e., garbage collectors. Automatic memory
management has received more attention from the verification community, with
numerous runtime systems and garbage collectors being studied and verified,
starting with \citet{doligez1993concurrent,doligez1994portable}, all the way to
CakeML's verified runtime system and GC~\cite{sandberg2019verified}. We
speculate that this is due to a proximity between the theorem proving community
and the garbage-collected, functional languages that they reason about and are
implemented in (e.g., OCaml). Alas, we do not see how to reuse the
foundational background of GC systems in the context of manual memory
allocation.

\section{Design and Overview of \name}
\label{sec:overview}

In this work, we design, implement and verify \name. Many of our design choices
leverage \hm~\cite{hardenedmalloc}, a modern, security-oriented memory allocator
heavily based on the OpenBSD malloc design. We therefore inherit many of \hm's
advantages, such as low memory fragmentation, low contention, a security-oriented default
design, and good long-term performance and scalability. To achieve these goals,
\hm, and therefore \name, rely on a variety of technical choices. We informally
review and describe these design choices here; \sref{spec} formally captures
those same design choices in \fstar.

\subsection{System Architecture}
\label{sec:sysarch}

\begin{figure}[t]
  \centering
  \includegraphics[width=\textwidth]{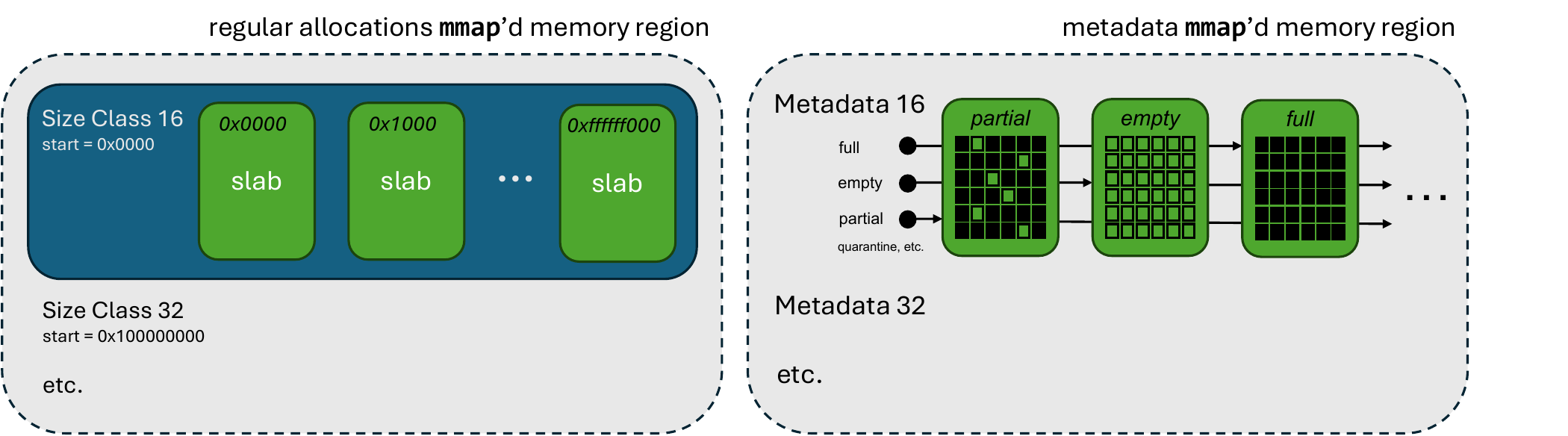}
  \caption{Architecture of \name.
  All regular allocations live in a single \li+mmap+'d memory region
  (left). All metadata lives in a single distinct \li+mmap+'d memory
  region (right). All allocations are contiguous. All metadata is contiguous.
  Accessing the n-th slab of the m-th size class thus boils down to
  pointer arithmetic, and similarly for the metadata. We rely on
  virtual memory to ensure only in-use pages occupy memory.
  Only a single arena is shown for conciseness;
  other arenas and their metadata are laid out contiguously in the same two
  \li+mmap+'d regions.}
  \label{fig:architecture}
\end{figure}

\fref{architecture} describes the overall architecture of \name. We proceed top-down, and
describe the sequence of steps taken in order to fulfill a memory allocation
request.

\myparagraph{AVL tree for large allocations}
If the allocation exceeds the page size (4kB), it is deemed a ``large
allocation''. Large allocations are tracked in an auxiliary map, implemented
using an AVL tree (not shown in \fref{architecture}). Large allocations are directly forwarded to the OS
(via \li+mmap+); once the syscall returns, \name receives a fresh address of the
desired size; inserts a new entry into the AVL tree that maps the fresh address
to its size; and returns the fresh address to the client. (We explain in the
\li+free+ paragraph, below, why keeping track of the large allocation size is
important.)

\myparagraph{Memory-mapped region for regular allocations}
If the allocation is equal to, or smaller than the page size (4kB), it is a
regular allocation. Relying on modern OSes and their
treatment of virtual memory, we perform a single call to \li+mmap+ at allocator
initialization-time, which reserves a region of virtual memory for
\emph{all} regular allocations. At the beginning, none of the corresponding
pages are used; unless the client program allocates substantial amounts of
memory, the virtual memory manager guarantees that the actual memory usage will
be correspondingly low. Throughout \name, we take care to return unused pages to
the OS to be used for other purposes, but retain the virtual address space which
remains unchanged throughout the execution of the program.

\myparagraph{Arenas}
The large memory region for regular allocations is first subdivided into
\emph{arenas}. The number of arenas is set at compile-time; using
thread-local storage, we associate an arena to each thread. Arenas are
\emph{not} protected by locks, which operate at a smaller level of granularity.
The point of arenas is to spread out allocations in order to statically reduce
contention. The allocator looks up the arena associated to the current thread,
then proceeds within this arena.

\myparagraph{Size classes}
Each arena is further subdivided into size classes, which contain allocations of a given size;
each regular allocation is rounded up to the nearest available size class.
Just like the number of arenas, the number of size classes is fixed:
there are 27 size classes for different multiples of 16 bytes up to the page size.
(We show later in this paper how \name is written for a
\emph{generic} number of size classes, which the user can trivially adjust at
\fstar compile-time. This paves the way for further extensions, e.g.,
handling ARM's 64kB large pages.)

Thanks to virtual memory, only the
portions of each size class that are in use are backed by physical memory.
Each size class is associated to metadata that needs to be updated whenever an
allocation takes place within that size class: for that reason,
size classes are each protected by an individual lock. Recall that the mapping of threads to
arenas is not injective (the number of arenas is fixed): the lock thus prevents
two threads from racing to allocate within the same size class.

\myparagraph{Slabs}
Each size class is made up of slabs. A slab is exactly one page, divided into
allocation units, called slots. All slabs within a size class are divided into the same
slot size. That is, within the 4kB size class, each slab contains one
slot; and within the 32B size class, each slab contains 128 slots.

Each size class tracks
all of its slabs using \emph{size class metadata}. The metadata keeps track of
free slabs, and to reduce memory fragmentation, it also separately tracks
partially filled slabs, which are prioritized for allocation. The tracking of
all three kinds of slabs is achieved using doubly-linked lists to efficiently
insert/remove when implementing malloc/free. Additionally, the size class
metadata contains another two extra lists for security (detailed in
\sref{spec-security}). (In practice, our lists have a special marker for the
last used slab, so as not to initially contain, e.g., 1M elements for the list of
free slabs.)

To fulfill an allocation request, the allocator acquires the lock for that size
class, looks for a partial of free slab (i.e., one where there is space left)
in the metadata, then proceeds.

\myparagraph{Slots}
Each slab is made up of individual slots. The design of \name ensures that
each individual allocation is aligned to 16 bytes, which guarantees better
performance on modern processors. To keep track of which slots are used or
available within a given slab, \name maintains \emph{slab metadata}; in
practice, this is a bitmap. We eschew free lists, because we cannot guarantee
that an attacker will be unable to corrupt them, even in the presence of an
encoding. 
Conversely, our bitmaps live in a separate
\li+mmap+'d region, which ensures that an attacker cannot materialize the
metadata virtual address.
This is the final step for the allocator: once a suitable slab is identified, a
slot is taken within that slab, the \emph{slab metadata} is updated, the size
class lock is released, and the address of the corresponding slot is returned to
the client program.

\myparagraph{Freeing}
In order to free, the allocator first performs a pointer
comparison to determine whether the pointer belongs to the mmap'd region of
regular allocations. If so, further pointer difference operations and bit
manipulations allow determining the arena and size class that the pointer
belongs to.
\name then
acquires the relevant lock, locates the suitable slab,
updates the bitmap (slab metadata), size class metadata, then releases the lock.
If this leads to a new empty slab, the corresponding slab is moved to quarantine
(more on quarantine below), and returned back to the OS by calling \li+mmap+ to
modify the flags for that page.
Should the pointer be invalid (i.e., not correspond to a slot), the allocator
leaves the bitmap unchanged and releases the lock.
If the pointer belongs to none of the allocator's arenas, then the pointer must
be a large allocation. A lookup in the AVL map allows i) ensuring that the
pointer is indeed valid, and ii) retrieving its size. This allows \name to call
\li+munmap+, which given a pointer and its size, releases the corresponding
pages back to the OS.

\myparagraph{Concurrency}
We mentioned above how our design relies on locks for size classes. We
note that this is a fine-grained concurrency design, aiming to mitigate
contention. A lock-free allocator would be ideal, but to the best of our
knowledge, cannot be implemented using segregated metadata, which is a crucial
security feature we are not willing to relinquish. Conversely, large-grained
concurrency at the level of arenas would be highly undesirable. Several threads
may be concurrently touching the same arena, but in different size classes;
fine-grained locks avoid needless contention on that arena. We remark that if,
even in the presence of our fine-grained locks, contention is still an issue,
the user can increase the number of arenas to better support highly threaded
programs.

\subsection{Allocator Hardening}
\label{sec:spec-security}

Our design not only incorporates all of the mechanisms described above, but adds
extra mechanisms in order to defend against heap corruption vulnerabilities.
This is consistent with our earlier claim that the allocator is the last line of
defense, but also highlights the need for verification: the design above is
non-trivial with many subtle invariants, and becomes even more so once the
security mechanisms are added.
Many hardening techniques have been, and continue to be proposed. We adopt the
most widely-implemented ones amongst security-oriented allocators~\cite{yun2021hardsheap},
suggesting their importance and effectiveness.
We now review
each one of the main security mechanisms \name implements.

\myparagraph{Segregated metadata}
As mentioned above, our metadata lives in a separate address space (i.e., an
\li+mmap+'d region that is disjoint from the regular allocation region).
If an attacker manages to
write past a given slot, the damage is limited since the allocator's internal
data structures are untouched.

\myparagraph{Zeroing}
Just like Apple's OSX allocator~\cite{apple-zero}, we zero-out memory upon freeing.
We do this either using a cryptographic-grade secure zeroing function (regular allocations)
which is resistant to rogue C compiler optimizations, or \ls`munmap` (large allocations).
This defends against an attacker observing sensitive data through a reused memory allocation slot.
Furthermore, upon allocation, we verify that the memory is still zeroed out.
If not, the memory was corrupted, and we return NULL as a further defensive security measure,
rather than returning possibly poisoned memory to the client program.

\myparagraph{Quarantine}
Once a slab becomes empty, we do not reuse it immediately and instead keep it in
quarantine. This is implemented using a fourth linked list within a given size
class, which keeps track of quarantined slabs. Unlike the other linked lists (full, empty, partial),
the quarantine list exposes a \emph{queue} interface (also implemented for \name).
The idea is to let the slab
``cool off'' to make any subsequent accesses to any pointer in it an OS
page fault error. This defends against double frees, and generally accessing
freed memory. In practice, we simply call \ls`madvise(DONT_NEED)` to return the page to the OS.
Using a queue allows us to reclaim pages from quarantine (so as
not to exhaust virtual memory tables) in a first-in, first-out order, relying on customizable
compile-time parameters to specify what is a sufficiently long time before a quarantined
page becomes reusable.

\myparagraph{Guard pages}
Within size classes, some of the slabs are marked as inaccessible via
\li+mmap+, meaning that any access to them causes a fault. We call these guard
pages. The chief reason for guard pages is to prevent against large buffer
overruns that may attempt to linearly read past an allocation unit. These are
tracked in the final, fifth doubly-linked list.

\myparagraph{Canaries}
Each slot is terminated by a canary, consisting of a magic number,
which can absorb small buffer overflows instead of corrupting
the next slot. We check the integrity of a canary
when freeing a slot, which can detect overflows \emph{a posteriori}
(but often too late to entirely prevent attacks).

\section{Modeling the Essence of a Security-Oriented Allocator}
\label{sec:spec}

\sref{overview} provided a high-level, informal description of \name.
We now show how to precisely and accurately capture the behavior of \name,
by specifying its well-formedness and functional correctness through the use
of a dependently typed concurrent separation logic.
We start by providing background about concurrent separation logic,
and its encoding into \fstar.

\subsection{Background}
\label{sec:background}
\myparagraph{Concurrent Separation Logic}
Separation logic~\cite{ohearn01seplogic,reynolds02seplogic,ishtiaq01bi}
is a Hoare-style program logic that is well-suited to reasoning about heap-manipulating
programs: the Hoare triple \hoare{P}{c}{Q} represents that, if precondition $P$ is initially
satisfied, then the program $c$ executes without error, that is, the program is memory safe,
and the postcondition $Q$ is satisfied when $c$ terminates, which can be used to specify
the functional correctness of the program.
As an example, the specification of memory allocation is \hoare{emp}{r =
alloc(v)}{\ptsto{r}{}{v}}, where $emp$ corresponds to the empty predicate,
always trivially true, while \ptsto{r}{}{v} represents that $r$ is live and
points to the value $v$.

Separation logic predicates can be composed: \estar{P}{Q} represents that $P$ and $Q$ relate
to separated (i.e., disjoint) regions of memory. This enables modular reasoning through the
separation logic frame rule: if \hoare{P}{c}{Q} holds, then so does 
\hoare{\estar{P}{R}}{c}{\estar{Q}{R}}; any predicate $R$ relating to a part of memory
separated from the execution of $c$ is left unchanged. 

Concurrent Separation Logic (CSL)~\cite{ohearn07csl, brookes07csl} extends separation logic
to provide guarantees in concurrent settings. In CSL, the Hoare triple
\hoare{P}{c}{Q} captures the thread-safety of the program $c$, namely,
the memory safety and functional correctness guarantees still hold in the presence
of threads executing concurrently.

\myparagraph{\fstar and \krml}
\fstar~\cite{mumon} is a proof assistant and program verification framework.
It allows users to write, specify, and prove programs in a functional language with a rich
type system based on dependent types and user-extensible effects~\cite{layeff}. To simplify
verification, \fstar discharges proof obligations using an automated SMT
solver~\cite{z3}, while also offering metaprogramming and tactic-based verification
facilities~\cite{metafstar} in a style similar to Lean~\cite{lean} or Idris~\cite{idris}.

While \fstar code extracts to OCaml by default, a subset of the \fstar language also
extracts to idiomatic C code through the \krml compiler~\cite{lowstar}.
This toolchain was previously successfully used to provide verified C implementations
of security-critical applications such as cryptographic
libraries~\cite{haclxn,evercrypt,ho2022noise} or binary parsers and
serializers~\cite{ramananandro2019everparse,swamy2022hardening}.

\myparagraph{Steel}
Steel~\cite{steelcore2020,steel2021} is a recent framework embedded in \fstar that allows
the use of concurrent separation logic to specify and prove the correctness and safety
of concurrent \fstar programs.
To introduce the reader to Steel and to the \fstar syntax
used throughout this paper, we present below the signature of the \ls`malloc` function
that we aim to implement.

\begin{lstlisting}
val malloc (size: size_t) : Steel (array uint8) emp (fun r -> null_or_slarray r)
  (requires fun  h0 -> True) (ensures fun  h0 r h1 -> not (is_null r) ==> length r >= size)
\end{lstlisting}

The \ls`malloc` function takes as argument the number of bytes \ls`size` that must
be allocated. It has the \ls`Steel` \emph{effect}, indicating that it belongs to the Steel
framework. This effect has several indices.
First, \ls`array uint8` corresponds to the
return type of this function.
The second and third indices are separation logic
predicates, of type \li+slprop+. They mimic the specification of memory
allocation we introduced earlier: \li+malloc+ consumes \li+emp+ (second index),
which carries no ownership, and returns
\ls`null_or_slarray r`, where \ls`r` refers to the
returned value. \ls`slarray r` is the Steel equivalent of \ptsto{r}{}{\_}, but applied
to arrays instead of references. To faithfully model \ls`malloc` according to the
C standard, we account for the fact that \ls`r` may be null.
Steel's separation logic predicates can capture
a rich class of properties about memory; for this paper's purposes, it is sufficient
to see them as representing a notion of memory ownership.
The fourth and fifth indices
lighten the formalism and simplify the verification
effort. In Steel, these allow us to specify properties about the contents of memory using the
\ls`requires` and \ls`ensures` clauses, operating on the initial memory \ls`h0` and
the final memory \ls`h1`. This style has been used successfully by tools such as
Dafny~\cite{dafny} or Steel's predecessor, Low$^\ast$~\cite{lowstar}, and seems
to be a natural style for writing these specifications. Here, the precondition
(fourth index) is trivial. The post-condition (fifth index) states that,
if the array returned is not \ls`NULL`, then its
length is at least the size initially requested.

When written in the \krml-compatible fragment, Steel code extracts to idiomatic C code.
For instance, the Steel \ls`malloc` signature above will yield the following standard
C function prototype: \li+uint8_t* malloc(size_t size)+. This signature is more precise
than that of the C standard
(\li+uint8_t*+ instead of \li+void*+); in practice, we add a wrapper that performs
a cast, so as to match the expected ABI and thus provide a drop-in replacement
for existing userspace memory allocators.
In the rest of this paper, we go over the various steps that ultimately produce
a memory allocator satisfying this Steel signature, thus guaranteeing
memory safety, functional correctness, and thread-safety.

\subsection{Specifying the Allocator Architecture}
\label{sec:allocspec}

We now present how to leverage dependently typed separation logic to
specify the architecture of \name.
Our specification is modular, and relies on several tiers of composable abstractions.
We give an overview from the ground up, starting with our handling of regular allocations.

\myparagraph{Tying Slots and Metadata}
As described in \secref{sec:sysarch},
a slab is a memory page seen as an array of slots of the same bytesize $s$.
During allocation, ownership of one of these slots is given to a client
when $s$ or less bytes of memory are requested. The role of the allocator is to keep
track of which slots were previously allocated, and which ones the allocator still owns, and
is therefore able to transfer to a client.
To do so, the allocator relies on \emph{slot metadata}.

Abstractly, the metadata \li+slots_md+ maps each slot index to a boolean \ls`b`,
indicating whether the slot is available for allocation (i.e., owned by the
allocator). Therefore, given a \li+slot+ and corresponding boolean \li+b+,
\li+available_slot b slot+ either states that the allocator owns the slot
(\li+b = true+, using the \li+slarray+ predicate we saw earlier), or states no
ownership at all.

\begin{lstlisting}
let available_slot b slot : slprop = if b then slarray slot else emp
\end{lstlisting}

Before stating ownership at the level of an entire slab, we take a detour and
introduce a helper combinator, called \li+starseq+, below, which given a
sequence \ls`s` of length \li+n+, and a predicate \li+p+, returns
\li+p s[0] 0 * ... * p s[n] n+. Unbound names
(e.g., the type \ls`a` below) are implicitly bound at the top.

\begin{lstlisting}
type idx (s:seq a) = i:nat{i < length s}
let starseq' (p: a -> nat -> slprop) (s:seq a) (i:idx s) = if i = length s then emp else p s.[i] i `star` starseq' p s (i + 1)
let starseq p s : slprop = starseq' p s 0
\end{lstlisting}

This combinator can be useful in a variety of contexts, not just for tying slots
to their metadata: its definition is fully generic and works for sequences of
any type \ls`a` with any separation logic predicate \ls`p`. To conveniently
reason about this predicate, we also provide generic lemmas to, e.g., extract
the \ls`slprop` corresponding to a given index, or to easily update the sequence
\ls`s` and modify \ls`starseq` accordingly without needing to unfold its
definition.
This programming style is representative of our methodology;
when possible, we define proof idioms and helpers generically, and instantiate them
when needed throughout the codebase. This reduces both code duplication and the proof
effort; we only need to perform proofs on the generic version, instantiations come
at no extra cost.

Our metadata is conceptually a map from slots to
booleans; in our specifications, however, we use a
smarter formulation that avoids tedious reasoning on the domain of the map.
Since we statically
know the number of slots in a slab, and that each slot has associated metadata,
we can represent metadata as a sequence of booleans of length equal
to the number of slots.
With that, we can define the ownership for an entire slab as
the conjunction of \ls`available_slot` for all
slots in the slab by iterating over all elements in the metadata sequence \ls`slots_md$_v$`,
using the \li+starseq+ predicate.

\begin{lstlisting}
type mempage = a:array uint8{length a == page_size}
let slots_sl size slots_md$_v$ (slab:mempage) : slprop =
  starseq (fun b i -> available_slot b slab[i*size..(i+1)*size-1]) slots_md$_v$
\end{lstlisting}

While we have been operating on a representation of metadata as a mathematical sequence,
metadata will in practice be a memory object (in our case, a bitmap),
that the allocator must be able to safely
access and update. Hence, we also need to specify that the allocator has ownership of
a memory region corresponding to well-formed metadata, and tie its contents to the
\ls`slots_sl` predicate above. We do it as follows, \emph{reflecting} the
current contents of the metadata \emph{as} a sequence of booleans via the
\li+Bitmap.as_seq+ helper:

\begin{lstlisting}
let slab_sl size (slots_md:bitmap) (slab:mempage) : slprop =
  sldep (bitmap_sl slots_md) (fun slots_md$_v$ -> slots_sl size (Bitmap.as_seq slots_md$_v$) slab) 
\end{lstlisting}

The \ls`sldep` predicate is a generic Steel predicate that captures the
combination of two \ls`slprop`s \ls`p` and \ls`q`,
where the second depends on the contents
of the first. In the example above, assuming that the bitmap \ls`slots_md` contains
the value \ls`slots_md$_v$`, the predicate \ls`slab_sl` thus corresponds to 
\ls+bitmap_sl slots_md `star` slots_sl size (Bitmap.as_seq slots_md$_v$) slab+. 

Separating into different composable abstractions is a key component
of our verification methodology. First, it enables modular verification,
where each component can be verified in isolation in a smaller proof context,
as we will see in \secref{sec:impl}. Second it simplifies code maintenance
and updates to the codebase. For instance, while we represent metadata as
a bitmap, switching to a different datastructure, e.g., a free list would
be straightforward, and only require changes to \ls`slab_sl`, and not
to the layers below.

\myparagraph{Slabs}
As described in \secref{sec:sysarch}, slab metadata keeps track of empty, partially filled,
and full slabs to efficiently find a slab in which to perform allocation while minimizing
memory fragmentation. Since clients can perform calls to malloc and free in (almost) any order,
slabs of different kinds can be intertwined. To efficiently access the next empty or partial
slab during allocation, the metadata relies on list-like structures.

While lists have a well-known, textbook separation logic specification, one difficulty arises
when attempting to use them to implement a memory allocator. List-like structures typically
rely on dynamic memory allocation when, e.g., creating and inserting a new cell, which
is exactly what our allocator must implement! To circumvent this circularity, we leverage
the fact that the maximum number of slabs is statically known
(based on the amount of memory available for the entire allocator), and instead implement
a library for linked lists where all cells are part of one single static array.
To do so, we first specify what is considered a well-formed list inside the
sequence \ls`slabs_md$_v$` as \ls`is_list`, below.
The predicate is mostly concerned with well-formedness of the sequence of indices,
i.e., they are all in bounds, and represent a non-cyclic list. In particular, this
predicate does not capture ownership, and has therefore type \ls`prop`, and not \ls`slprop`.

\begin{lstlisting}
type cell (a:Type) = {data : a; next: size_t}
let rec is_list' (slabs_md$_v$:seq (cell a)) (idx:size_t) (count:nat) : prop =
  if count > length slabs_md$_v$ then False              // Cyclic list 
  else if idx = null_idx then True                         // Terminal case
  else if idx < 0 || idx >= length slabs_md$_v$ then False // Out-of-bound
  else is_list' slabs_md$_v$ (slabs_md$_v$[idx].next) (count+1)
let is_list slabs_md$_v$ idx = is_list' slabs_md$_v$ idx 0
\end{lstlisting}

Equipped with this, we can now define the separation logic predicate capturing
ownership of three well-formed linked lists with head pointers \ls`idx$_e$`, \ls`idx$_p$`,
and \ls`idx$_f$` (for empty, partial, and full slabs respectively) inside the array \ls`ptr`.
To do so, we use \ls`slrefine`, another generic Steel separation logic combinator.
\ls`slrefine slp p` captures the ownership of the separation logic predicate \ls`slp`,
and that \ls`p` holds on the memory contents of \ls`slp`.
In addition to the well-formedness of the lists, we also require that they 
partition all of the slabs metadata: all slabs are indeed managed by the
allocator.
\begin{lstlisting}
let arraylist_sl (ptr:array (cell a)) (idx$_e$ idx$_p$ idx$_f$: size_t) : slprop = slrefine (slarray ptr) (fun slabs_md$_v$ ->
  is_list slabs_md$_v$ idx$_e$ /\ is_list slabs_md$_v$ idx$_p$ /\ is_list slabs_md$_v$ idx$_f$ /\ partition slabs_md$_v$ idx$_e$ idx$_p$ idx$_f$)
\end{lstlisting}

Since the lists dynamically evolve, the head pointers can change, e.g.,
when inserting at the head of the list. We thus must store their
current values in valid memory locations, on which the allocator must have ownership.
This can be captured by reusing the \ls`sldep` predicate seen previously.
\begin{lstlisting}
let ind_arraylist_sl (r: array (cell a)) (r$_e$ r$_p$ r$_f$: ref size_t) =
  sldep (ptr r$_e$ `star` ptr r$_p$ `star` ptr r$_f$) (fun (idx$_e$, idx$_p$, idx$_f$) -> arraylist_sl r idx$_e$ idx$_p$ idx$_f$)
\end{lstlisting}

Just like before, we briefly comment our design choices. Here, because of the
fixed size of our allocations, our ``arraylist'' is a much simpler separation
logic predicate than an actual linked list, which would require us to follow
series of pointers through the heap, and would also require us to prove
their termination. We avoid this specification burden and we simply talk about
indices within statically-known bounds, in our opinion a much more judicious
choice.

The last step is now to instantiate this generic linked list with the
concrete metadata for slabs, and to tie it to the underlying memory and slots metadata.
In our case, slab metadata will correspond
to a \ls`status`, represented as an enumeration of the possible slab states.
In memory, the slab metadata will therefore be an array of \ls`cell status`,
which permits efficient accesses and updates of metadata; if the list structure
does not need to be modified, these operations are performed in constant time.
\begin{lstlisting}
type status = | Empty | Partial | Full
let empty_slab md$_v$ = forall (i:nat{i < length md$_v$}). md$_v$.[i] == true
let dispatch stat (slab_md:bitmap) (slab:mempage) = slrefine (slab_sl slab_md slab) (fun (md$_v$, _) ->
  match stat with | Empty -> empty_slab md$_v$ | Partial -> partial_slab md$_v$ | Full -> full_slab md$_v$)

let slabs_sl r$_e$ r$_p$ r$_f$ (slabs_md:array (cell status)) (slots_md:array bitmap) (mem:array uint8) = sldep
  (ind_arraylist_sl slabs_md r$_e$ r$_p$ r$_f$) (fun (idx$_e$, idx$_p$, idx$_f$, slabs_md$_v$) ->
    starseq (fun stat i -> dispatch stat (ith_slot slots_md i) (ith_page mem i)) slabs_md$_v$)
\end{lstlisting}

This specification combines several of the generic predicates seen previously.
First, for a given slab, \ls`dispatch` refines \ls`slab_sl`
with additional constraints linking the slab status to its contents. For an \ls`Empty` slab,
all slots in the slab must be free, as captured by \ls`empty_slab`.
Predicates for partial and full slabs (omitted) are similar.
To generalize this property to all slabs, we combine \ls`dispatch` with
\ls`starseq`; \ls`ith_slot` and \ls`ith_page` (omitted) select
the subarrays corresponding to the i-th slot metadata and i-th memory page respectively.
Finally, we tie the memory and slot metadata to the slab metadata using the \ls`sldep`
combinator; this captures that they are in separated, independent regions of memory.

The use of our generic combinators to structure specifications here has several advantages;
first, they allow us to better separate different concepts, both in the specification
and in the later proof that the implementation abides by it. For instance, we can reason
separately about the shape of the slab metadata (i.e., linked lists in a static array),
and its contents, which determine the structure and availability of the underlying slots.
This enables easier to read specifications, where irrelevant parts in a given context can
remain abstract. Additionally, this permits writing modular, verified code by following
the specification which simplifies code maintenance. As an example, during the development
of \name, we replaced the linked lists presented in this section by
doubly-linked lists to improve performance. 
This change only required replacing the definitions of \ls`arraylist_sl` and providing
functions to perform list operations with the same interface as \ls`arraylist`. The code
and proofs operating on the contents of the lists (i.e., the slabs metadata) was left
unchanged.

\myparagraph{Size Classes}
Now that we captured the memory layout of and interaction between data, slab metadata,
and slots metadata, all that remains is to gather slabs into size classes for allocation
units of different sizes. To do so, we first collect all relevant values for a given
size class into a struct, of type \ls`sc_data`. The size of memory handled by a size
class, called \ls`sc_size`, is fixed statically and is defined as the size of a memory
page (given by the OS) multiplied with a user-provided number of pages per size class.
The \ls`size` field represents the size of an allocation unit for this size class;
we restrict the values it can take to respect alignment constraints.

\begin{lstlisting}
type sc_data = {
  size: size_t{size % 16 == 0 /\ size > 0};
  empty_ptr: ref size_t;
  partial_ptr: ref size_t;
  full_ptr: ref size_t;
  mem: array uint8{length mem == sc_size};
  slabs_md: array (cell status){length slabs_md == nbr_pages};
  slots_md: array bitmap{length slots_md == nbr_pages}; }
\end{lstlisting}

Building upon the previous predicates, defining the predicate \ls`size_class_sl`
corresponding to one size class is now straightforward.
With these definitions in hand, we can now provide the following Steel signature
for allocation inside one size class.
\begin{lstlisting}
let size_class_sl (sc: sc_data) = slabs_sl sc.empty_ptr sc.partial_ptr sc.full_ptr sc.slabs_md sc.slots_md sc.mem
val sc_malloc (sc: sc_data) : Steel (array U8.t) (size_class_sl sc) (fun r -> size_class_sl sc `star` null_or_slarray r)
  (requires fun h0 -> True) (ensures fun h0 r h1 -> not (is_null r) ==> length r == sc.size)
\end{lstlisting}

\myparagraph{Alignment}
In C, memory accesses where the address is not a multiple of the size of the object being accessed
are undefined behavior. To make things tractable, the C standard thus dictates that the values
returned by \li+malloc+ be suitably aligned for ``any object with a fundamental alignment
requirement''. In practice, this means that \li+malloc+ satisfies the alignment requirements of the
largest scalar type, typically \li+long double+, which is either 8 or 16 depending on the ABI.
Naturally, larger alignments are oftentimes needed, for instance when using vector types (SSE, SSE2). To
that end, C11 and later define \li+aligned_alloc+, which supersedes
previous extensions such as \li+memalign+ and \li+posix_memalign+. All the standard specifies is
that \li+aligned_alloc+ may fail if the alignment is invalid or greater than that supported by the
platform.

We implement both regular (\li+malloc+) and aligned (\li+aligned_alloc+) APIs, along with wrappers
that provide ABI compatibility with non-standard APIs that predate C11. As to our implementation
choices, we follow typical behavior and support any power of two between 16 and
the page size. We could easily support larger alignments, but have not yet found a reason to do so.
For simplicity and performance, our \li+malloc+ returns values that are always aligned on a 16-byte
boundary, which is stricter than what some older ABIs (e.g., Linux x86) permit.
To achieve this, we carry around a predicate \li+aligned+ that appears at slab-allocation time,
namely, \li+aligned page_size+. This captures our assumption that calls to \li+mmap+ always return
page-aligned memory. The predicate then makes its way throughout our code, and is eventually
provided to the client as a post-condition of \li+malloc+ (\li+aligned 16+) or \li+aligned_malloc+
(\li+aligned n+). To facilitate reasoning within \name, we mostly keep the predicate abstract.
Non-linear arithmetic is circumscribed to one well-controlled lemma that goes from
\li+aligned page_size+ to smaller alignment constraints, thus limiting issues with SMT solvers.

\myparagraph{Concurrency}
The function \ls`sc_malloc` above
is close to our target signature for \ls`malloc` outlined in
\ref{sec:background}, with one key difference: it requires (and returns after execution)
exclusive ownership of the size class through \ls`size_class_sl`. 
When operating in a concurrent setting, guaranteeing ownership can be tricky; we must ensure
that no thread executing concurrently also accesses the size class.

To do so, we rely on locks, which have a natural model in CSL.
A CSL lock is a value associated to a given predicate, with
(blocking) \ls`acquire` and \ls`release` functions, shown below. We refer
the interested reader to~\cite{steel2021,jung2018iris} for a more detailed
presentation of CSL locks.
\begin{lstlisting}
val lock (p:slprop) : Type
val acquire (l: lock p) : Steel unit emp (fun _ -> p)
val release (l: lock p) : Steel unit p (fun _ -> emp)
\end{lstlisting}

We leverage this to provide fine-grained concurrency, where, instead of relying on one
global lock, we associate one lock to each size class, packed together in one dependently
typed struct.

\begin{lstlisting}
type size_class = {
    data : sc_data; // The content of the sizeclass
    lock : lock (size_class_sl data); // Mutex locking ownership of the sizeclass }
\end{lstlisting}

We store the different size classes inside a global array, shown below. However,
as the array resides in memory, we now need to retrieve its ownership before accessing it.
Hiding this ownership behind a global lock would not be acceptable,
as it would make our fine-grained concurrency design
useless. Instead, we observe that, after initialization,
all accesses to the size classes
are read-only; the specificities of size classes are fixed at compile-time.
To avoid relying on locks, we instead implement a novel model of
\emph{frozen arrays} in Steel, that can be safely accessed concurrently
by any thread once initialized, but do not provide any update facility.

\begin{lstlisting}
val scs : frozen_array size_class{length scs == nb_sc}
\end{lstlisting}

We finally show the entrypoint of our allocator for regular allocations,
the \ls`small_malloc` function,
with the signature outlined in \secref{sec:background}. For presentation purposes, we assume
that our allocator only contains two size classes, we will show in
\secref{sec:meta} how to make their exact number configurable in our implementation.
In practice, to further reduce contention, we also separate size classes into different arenas;
the global \ls`size_classes` frozen array becomes of size \ls`nb_sc * nb_arenas`; if
\ls`small_malloc` is called by a thread with arena id \ls`i`, it only tries to allocate
in the size classes between indices \ls`i * nb_arenas` and \ls`(i + 1) * nb_arenas` excluded.

\begin{lstlisting}
inline_for_extraction with_lock lock f = acquire lock; let res = f () in release lock; res
let small_malloc bytes =
  let sc = scs[0] in if bytes <= sc.data.size then with_lock sc.lock (fun () -> sc_malloc sc.data)
  else let sc = scs[1] in if bytes <= sc.data.size then with_lock sc.lock (fun () -> sc_malloc sc.data)
  else return null
\end{lstlisting} 

One important thing to note is that,
while CSL and CSL locks ensure thread-safety, this model does not allow to formally prove the absence
of deadlocks. To alleviate this issue, we syntactically ensure that all uses of locks are done
in a scoped way. To do so, we define a combinator, called \ls`with_lock`, which takes as arguments
a lock and a thunked Steel function. This combinator is annotated with the \fstar \ls`inline_for_extraction`
attribute, which indicates to unfold its definition at compile-time; in the generated C code all calls
to \ls`with_lock` are replaced by the body of the combinator. 

\myparagraph{Large Allocations}
Compared to regular allocations, the specification and implementation of large allocations
is much simpler. Large allocations are performed through direct calls to \ls`mmap`; the allocator
keeps track of the pointers that have been allocated using a map from pointers to their length
to free the correct amount of data through \ls`munmap`.

We adopt a similar generic methodology for large allocations; the specification and implementation
rely on an abstract model of a map and of its basic primitives (e.g., insertion). We then
instantiate this abstract model with a new implementation of AVL trees in Steel; using any
other data structure would be straightforward. Similarly to size classes, we rely on a lock
to ensure the thread-safety of large allocations. A first version of our implementation called
\ls`mmap` at each AVL node allocation, which was highly inefficient; to improve the performance of large allocations,
we instead reused the slab allocator previously presented. Thanks to
our tight abstractions and abstract model, performing this change was straightforward. 

\subsection{Specification of Security Mechanisms}
While the presentation so far has focused on functional properties of the allocator,
we now present how to extend our formalization to support a range of security mechanisms.
Leveraging the modular abstractions previously presented, we show how changes to the
existing code are concise, and do not require large-scale modification across the code base.
Our chief result is that the allocator remains correct and memory-safe in the presence
of all these security mechanisms; proving that the implementation of these mechanisms
effectively defend against attacks would require an attacker model that is,
to the best of our knowledge, beyond what current state-of-the-art in program
verification can do. That being said, we certainly hope to tackle this in future
work.

\myparagraph{Zeroing on Allocations}
To specify that allocations must be zero-filled, we extend the
specification of \ls`malloc` presented in \secref{sec:background}
as follows, adding an extra postcondition in the \ls`ensures` clause.

\begin{lstlisting}
val malloc (size: size_t) : Steel (array uint8) emp (fun r -> null_or_slarray r) (requires fun _ -> True)
  (ensures fun _ r h1 -> not (is_null r) ==> (length r >= size /\ (forall (i:nat{i < size}). get (h1.[r]) i == 0)))
\end{lstlisting}

The postcondition ensures that the contents
of the returned pointer \ls`r` in the final state \ls`h1` only consists of
zeros.
As mentioned in \sref{sysarch}, our free operation takes care of zeroing,
and our malloc operation (and others) takes care of checking that memory is still zero.
If the memory is corrupted, the allocator returns NULL, thus satisfying the postcondition.

\myparagraph{Quarantine}
Quarantined slabs are a special kind of slabs, disjoint from the full, partial or empty
slabs previously presented. To keep track of which slabs are quarantined, we
extend the slab metadata to support a fourth doubly linked list, associated to a fourth
status. 

\begin{lstlisting}
type status = Empty | Partial | Full | Quarantine
let dispatch stat slab_md slab = slrefine (slab_sl slab_md slab) (fun (md$_v$, _) ->
  match stat with  ... | Quarantine -> quarantine_slab md$_v$)

let arraylist_sl ptr idx$_e$ idx$_p$ idx$_f$ idx$_q$ = slrefine (slarray r) (fun md$_v$ -> ... /\ is_list md$_v$ idx$_q$)
\end{lstlisting}

To enable the reuse of quarantined slabs, we define a queue data structure on
top of doubly linked lists. This permits to easily access the first quarantined slab,
and unquarantine it by moving it to the list of empty slabs after a sufficiently
long time; we omit this for presentation purposes.

Thanks to our modular approach coupled with strong abstractions, required changes
to the codebase are highly localized, and most
of our code is left unchanged. In addition to the changes shown above, we need
to add a fourth head pointer to the \ls`sc_data` structure, and modify the code
and proof relating to the \ls`slabs_sl` predicates, but all layers above (e.g.,
operating on top of \ls`sc_malloc`) or below (e.g., code operating at the slots
level, on the \ls`slab_sl` predicate) are entirely unchanged.

\myparagraph{Guard Pages}
Our specification extension to support guard pages is similar to the one previously
presented for quarantine: we add a fifth linked list to the slabs metadata,
and perform localized changes in the codebase accordingly.
One key difference however is that the separation logic predicate we devise,
\ls`slab_guard`, is an abstract predicate for which we only
provide an introduction rule, as shown below. This induces a form of monotonicity: once
a slab is marked as \ls`slab_guard`, by virtue of abstraction, it can never be eliminated
to retrieve a separation logic predicate on which the allocator could operate, or that
it could transform to an \ls`slarray` to pass to the client.
Under the hood, the introduction rule is implemented as a call
to \ls`mmap` with the \ls`PROT_NONE` flag; this ensures that any access to a guard page
will result in a \ls`SIGSEGV` returned by the OS.

\begin{lstlisting}
let dispatch stat slab_md slab = match stat with ... | Guard -> slab_guard slab
val slab_guard slab : slprop
val slab_guard_intro slab : Steel unit (slarray slab) (fun _ -> slab_guard slab)
\end{lstlisting}

Additionally, guard pages must appear following a specified pattern; typically,
every \ls`guard_int` pages, where \ls`guard_int` is a statically
fixed, user-defined constant. To specify this, we extend the \ls`slabs_sl` predicate,
and in particular its \ls`starseq` component iterating over all slabs as follows.
\ls`guard_int` and \ls`guard_start` are two global Steel constants that
can be easily changed without modifying other parts of the code, thus making
experimenting with different guard pages patterns trivial.

\begin{lstlisting}
let slabs_sl ... = ... starseq (fun stat i -> dispatch ... /\ (i % guard_int = guard_start) ==> stat = Guard) ...
\end{lstlisting}

\section{Implementing, Verifying, Extracting to C}
\label{sec:impl}

We mentioned earlier that Steel admits a subset that compiles to C. We now
implement our allocator in that subset; show that it \emph{refines} the
specification from \sref{spec}; and demonstrate how to extract it to C while
keeping the implementation generic.
We thus obtain a provably correct implementation of the security-oriented \name, which
currently consists of 26,000 lines of \fstar code,
extracting to about 4,300 lines of executable C code. Our code is open-source on GitHub and,
to the best of our knowledge, is the second-largest Steel project to date.

\myparagraph{Specification-Guided Implementation}
To enable efficient verification, we adopt a specification-guided (also called \emph{proof-oriented})
approach when implementing \name. Concretely, our implementation closely follows
the abstractions and structure of the specification presented in \secref{sec:spec}.
This yields several benefits. First, particularly complex pieces of code (e.g.,
optimized implementation of bitmaps) are naturally circumscribed within the
boundaries of the specification-driven abstractions.
Second, verification conditions to prove that the code satisfies its functional
specification are simpler, as we leverage the successive layers of abstractions
to hide as many implementation details as possible; this greatly reduces the
proof effort needed to establish the safety and functional correctness of our
allocator.
Third, by leveraging the modularity of the specification and \fstar's
abstraction facilities, our code admits a natural separation into different modules
(e.g., to operate at the slots level or at the slabs level) where the implementation
details can be hidden behind strong abstractions (interfaces in \fstar jargon).
As mentioned in \secref{sec:allocspec}, this allows to easily perform code
optimizations (e.g., by replacing linked lists by doubly linked lists) without impacting
the rest of the code and proofs, as long as the interfaces are left unchanged.
This also permits highly parallel builds, which greatly reduce the
verification time and the development cycles: if module A depends on module B,
\fstar only needs to verify the \emph{interface} of module B, not its complete
implementation before verifying module A. Case in point, the entire verification
of \name only requires 30-40 minutes on a modern laptop;
in practice, we rarely perform builds from scratch during a development cycle, making
iterating much faster.

\myparagraph{Static Parametrization, Partial Evaluation}
\label{sec:meta}
General-purpose allocators such as \name can be used in many contexts;
thus, they must be easily configurable. Clients operating 
in a purely sequential setting might want do disable arenas to avoid memory waste,
while clients that only perform very small allocations might want to
disable the larger size classes.
When modifying the allocator configuration, some changes simply require modifying
global variables (e.g., to adapt the size of a global array), but others are more
involved, requiring deeper changes to the allocator's implementation (e.g.,
to initialize different numbers of size classes with different shapes).

When implementing an unverified allocator in C, a common technique is
to rely on preprocessor macros to generate suitable code.
This is however error-prone, and not ideal for security-critical
code. Performing deep modifications into the code is also not acceptable;
even in a verified setting where these changes could be safely performed,
the effort involved would deter most clients.

We leverage \fstar's normalizer, a key component of dependently typed languages
which allows reducing pure terms at \fstar-compile time, that is, before
extraction to C takes places -- one can think of it as a principled preprocessor that is
even more powerful that C++ templates.
We use the normalizer to verifiably specialize our code
according to a user-provided configuration.
Following the methodology outlined in the previous sections, we implement and verify our
allocator in a generic style, abstracting over the configuration details.
For instance, size class initialization is performed through the recursive
function \ls`init_size_classes` below, which takes as arguments arrays
for the size classes and metadata, as well as a list of sizes \ls`l` representing 
the size classes configuration.

\begin{lstlisting}
let rec init_size_classes size_classes md l = match l with
  | [] -> ()
  | slot_size :: tl -> init_size_class slot_size size_classes[0] md[0];
                 init_size_classes tl size_classes[1..] md[1..]
let init size_list =
  let size_classes = mmap .. in let md = mmap .. in
  init_size_classes size_classes md size_list
\end{lstlisting}

While this code is valid in Steel and successfully verifies, it is however not compatible with
\krml extraction: mathematical lists cannot natively extract to C.
For presentation purposes, we will assume that a client wishes to have
two size classes, of slot size 16B and 32B respectively; they will specify this by modifying
the global variable \ls`sc_list`. We rely on the \fstar primitive \ls`norm` to instruct
the compiler to reduce the application \ls`init sc_list`, according to the reduction steps
\ls`custom_steps` (omitted), which include for instance unfolding
the definition \ls`sc_list` and reducing pattern matchings operating on lists, but
not unfolding the definition \ls`init_size_class`. 
After normalization, all occurrences of lists disappeared, which
enables \krml to extract to the idiomatic C \ls`init` shown below.

\begin{lstlisting}
let sc_list = [16; 32]
let init = norm custom_steps (init sc_list)

void init() { // Generated C code after normalization
  size_classes = mmap .. ;
  md = mmap ..;
  init_size_classes size_classes[0] md[0] 16;
  init_size_classes size_classes[1] md[1] 32; }
\end{lstlisting}

This approach makes it straightforward to configure \name according to a client's needs.
We only need to modify the value \ls`sc_list` and to recompile our project; apart from
this variable, all the code and proofs are left unchanged. We apply this methodology
to all parts of \name where the static configuration impacts the shape of the code;
this includes for instance selecting an appropriate size class in which to allocate,
and handling arenas. We emphasize that this specialization at extraction-time is entirely
verified, and hence, untrusted; \fstar's normalizer is a key component of its typechecker,
and hence already part of the TCB.

\myparagraph{Generic, Reusable Libraries}
As part of the development of \name, we defined and implemented a range of generic libraries
which we instantiated to fit our needs. We already presented several of them
in \secref{sec:spec}; they include datastructures (statically allocated doubly linked lists,
efficient bitmaps, AVL trees, frozen arrays, FIFO queues), proof idioms and generic separation logic combinators
(\ls`starseq`, extensions and helpers for \ls`slrefine` and \ls`sldep`),
concurrency primitives (separation logic locks with native extraction to pthread mutexes),
and modeling of the interaction with the OS
(axiomatization of memory-related syscalls, e.g., \ls`mmap`).
These libraries required significant engineering and modeling effort, and
represent a large
part of our development, totalling close to 13,000 lines of \fstar code.
While crucial to the development of \name, they are a contribution of their own,
which significantly decreases the effort required for future systems
verification projects in \fstar. We organized our codebase to ensure they are entirely separate
from the core of \name, and hope to upstream them to the Steel standard library in the future to
benefit to the community.

Native, idiomatic extraction of locks was contributed to \krml for
this work, and will support further \fstar concurrent verification developments.
This required reconciling the Steel model of locks (where locks are values)
with a low-level extraction that relies on pthread and its semantics (where
the address of a lock matters). We achieved this by adding dedicated
passes in the \krml compiler, allowing us to support global
variables that contain locks (the static array of size classes). This represents
a significant improvement over the pre-existing implementation of locks that
relied on a simple compare-and-swap primitive.

\myparagraph{Other Supported Features}
While our presentation focused on \ls`malloc` for brevity reasons, applications typically
rely on all memory management primitives specified by the C standard~\cite[\S7.20.3]{c99}.
To be usable in production by real-world applications, \name thus also specifies and implements
the other primitives specified by the C99 standard, i.e., \ls`free`, \ls`realloc`, and \ls`calloc`.
It also includes recent extensions, such as \ls`aligned_alloc` introduced in the C11
standard~\cite[\S7.22.3]{c11}.

\section{Evaluation}

We now compare \name with other state-of-the-art memory allocators, including
security-oriented allocators, such as
\hm~\cite{hardenedmalloc},
DieHarder~\cite{novark2010dieharder}, and regular allocators, such as
the glibc and the next-gen musl allocators.

Doing so, we aim to i) evaluate the performance of \name against other
allocators;
ii) determine whether \name can be used as a drop-in replacement
in existing, widely used applications and iii)
assess the time and effort spent
developing \name, and whether the methodology can be expanded to support
further optimizations (security or performance) with reasonable time and effort.

Our evaluation relies on two sets of benchmarks. To provide a fine-grained performance comparison
with other allocators, we reuse mimalloc-bench~\cite{mimallocbench}, a comprehensive
and popular suite
which contains a selection of benchmarks from the academic
literature and several real-world applications. 
To evaluate the real-world readiness of our allocator,
we finally describe our integration of \name within Firefox,
which required an extensive set of allocator features. 

\subsection{Performance Evaluation}
\label{sec:perf-eval}
We start by evaluating the performance of \name, comparing the execution time and memory usage.
Throughout this work, we improved mimalloc-bench by allowing the use of the testsuite
with recent compiler toolchains and fixed several compilation issues and bugs in benchmarking scripts;
we contributed several patches that were merged into the mimalloc-bench repository.

\myparagraph{Experimental setup}
We run experiments on a 4-core, 8-thread machine with an Intel(R)
Xeon(R) CPU E5-1620 0 @ 3.60GHz processor and 12GB memory, using Linux 6.1 and glibc 2.38.
To ensure reproducible results, we use the ``performance'' scaling governor
(meaning all cores pegged at maximum speed without dynamic CPU scaling),
and increase \texttt{vm.max\_map\_count} from 65,530 to 1,048,576 to ensure
allocators using a large number of memory mappings, such as \hm and \name for
guard pages, work properly.

\myparagraph{Total Execution Time}
As we mentioned before, a meaningful way to measure an allocator's performance
is to measure the client program's total execution time. We evaluate several
state-of-the-art allocators on a variety of realistic workloads from mainstream
programs; complete results are available in the Appendix.
We focus on
the most salient comparisons in \tref{perf-rss}, where we consider \name, along with the
allocators mentioned earlier.
The time difference between \name and \hm measures the cost of verification, or more precisely,
the penalty for the last few performance optimizations we have yet to implement. The
time difference between \hm and glibc measures the cost of security: while performant
and widely-used, glibc's allocator design is known to be a possible liability from
a security perspective.

\begin{table}
  \centering
  \small
  \begin{tabular}{l|rrrrr|rrrrr}
   & \multicolumn{5}{c|}{Time} & \multicolumn{5}{c}{RSS} \\
  \hline
  Benchmark & dh & hm & mng & st & glibc & dh & hm & mng & st & glibc\\ \hline
  barnes & 1.036 & 1.0 & 1.0 & 1.0 & 1.0 & 0.774 & 1.0 & 0.745 & 0.832 & 0.699\\
  cfrac & 1.446 & 1.0 & 0.738 & 1.278 & 0.544 & 0.261 & 1.0 & 0.1 & 0.81 & 0.113\\
  espresso & 1.273 & 1.0 & 1.301 & 2.284 & 0.707 & 0.18 & 1.0 & 0.066 & 1.399 & 0.062\\
  gs & 2.278 & 1.0 & 0.678 & 1.304 & 0.339 & 0.797 & 1.0 & 0.524 & 1.39 & 0.513\\
  larsonN & 17.097 & 1.0 & 2.251 & 1.036 & 0.09 & 0.659 & 1.0 & 0.369 & 2.277 & 0.664\\
  larsonN-sized & 16.823 & 1.0 & 2.2 & 1.008 & 0.088 & 0.69 & 1.0 & 0.351 & 2.292 & 0.78\\
  leanN & 6.249 & 1.0 & 1.421 & 1.086 & 0.617 & 1.421 & 1.0 & 0.688 & 1.417 & 0.78\\
  lua & timeout & 1.0 & 1.05 & 1.554 & 0.729 & timeout & 1.0 & 0.639 & 1.338 & 0.654\\
  mathlib & ERR & 1.0 & 0.779 & 1.221 & 0.645 & ERR & 1.0 & 0.626 & 2.004 & 0.676\\
  redis & 2.809 & 1.0 & 0.82 & 16.379 & 0.592 & 0.493 & 1.0 & 0.216 & 1.339 & 0.205\\
  rocksdb & ERR & 1.0 & ERR & 0.843 & 0.813 & ERR & 1.0 & ERR & 1.471 & 0.785\\
  z3 & 5.063 & 1.0 & 0.813 & 1.375 & 0.688 & 0.622 & 1.0 & 0.417 & 1.294 & 0.401\\
  \hline
  alloc-test1 & 2.051 & 1.0 & 0.922 & 1.291 & 0.515 & 0.447 & 1.0 & 0.292 & 1.471 & 0.305\\
  alloc-testN & 12.566 & 1.0 & 3.693 & 1.574 & 0.136 & 0.302 & 1.0 & 0.247 & 2.678 & 0.201\\
  cache-scratch1 & 1.006 & 1.0 & 0.994 & 0.994 & 0.994 & 0.205 & 1.0 & 0.133 & 0.445 & 0.133\\
  cache-scratchN & 0.977 & 1.0 & 0.977 & 1.0 & 0.977 & 0.22 & 1.0 & 0.144 & 0.555 & 0.159\\
  cache-thrash1 & 1.006 & 1.0 & 0.994 & 1.0 & 0.994 & 0.219 & 1.0 & 0.143 & 0.401 & 0.143\\
  cache-thrashN & 1.0 & 1.0 & 1.0 & 1.0 & 1.0 & 0.204 & 1.0 & 0.135 & 0.517 & 0.133\\
  glibc-simple & 1.5 & 1.0 & 1.187 & 1.931 & 0.453 & 0.248 & 1.0 & 0.055 & 0.744 & 0.058\\
  glibc-thread & 10.033 & 1.0 & 3.161 & 1.679 & 0.074 & 0.133 & 1.0 & 0.032 & 2.813 & 0.056\\
  malloc-large & 3.604 & 1.0 & 1.034 & 1.046 & 0.539 & 0.928 & 1.0 & 0.918 & 0.928 & 1.196\\
  mleak10 & 1.692 & 1.0 & 0.923 & 1.0 & 1.231 & 0.284 & 1.0 & 0.07 & 0.682 & 0.146\\
  mleak100 & 1.754 & 1.0 & 1.0 & 1.041 & 1.369 & 0.258 & 1.0 & 0.062 & 0.618 & 0.128\\
  mstressN & 5.251 & 1.0 & 1.722 & 1.605 & 0.435 & 0.811 & 1.0 & 0.681 & 1.12 & 1.521\\
  rbstress1 & 1.14 & 1.0 & 0.809 & 0.895 & 0.812 & 1.177 & 1.0 & 0.905 & 0.925 & 0.708\\
  rbstressN & 1.285 & 1.0 & 0.782 & 0.939 & 0.76 & 0.705 & 1.0 & 0.684 & 0.941 & 1.0\\
  rptestN & 7.521 & 1.0 & 1.414 & 19.537 & 0.387 & 0.534 & 1.0 & 0.234 & 2.59 & 0.254\\
  sh6benchN & 22.109 & 1.0 & 3.239 & 1.649 & 0.175 & 1.771 & 1.0 & 0.868 & 1.12 & 1.056\\
  sh8benchN & 13.098 & 1.0 & 2.229 & 1.256 & 0.133 & 1.075 & 1.0 & 0.803 & 1.21 & 0.982\\
  xmalloc-testN & 8.978 & 1.0 & 1.543 & 1.17 & 0.205 & 0.836 & 1.0 & 0.066 & 0.393 & 0.815\\
  \end{tabular}
  \caption{Execution time and peak RSS for \name (``st'', this work),
  \hm (``hm'', \cite{hardenedmalloc}),
  DieHarder~\cite{novark2010dieharder} (``dh''),
  nextgen implementation for musl libc (``mng'', \cite{muslmalloc}),
  and the glibc allocator (``glibc'').
  Results are normalized and presented relative to \hm. ERR corresponds to crashes during execution.}
  \label{tab:perf-rss}
\vspace{-1cm}
\end{table}

Recall that our allocator is generic and that numerous parameters can be
tweaked. Aiming for an apples-to-apples comparison, we use
the same configuration as \hm, namely, the guard page interval set to 2
and the number of arenas set to 4, with quarantine and zeroing enabled.

The most meaningful comparison is thus between \name and \hm. The first series
of benchmarks are real-world applications.
For most applications, the performance overhead is within 0.84x to 1.38x; 
one notable outlier is \li+redis+, a key-value database. After careful inspection of the code,
differences in \li+redis+ seem to be due to a large number of allocations
of size slightly exceeding the page size. While these allocations
fall into \name's large allocations, requiring repeated calls to \ls`mmap`,
\hm's slab allocator can be used for allocations
of up to 131072 bytes (i.e. 32 standard 4k pages) by combining several pages together.
We did not yet implement this optimization in \name,
but do not foresee any difficulty doing so. We remark that compared to another defensive allocator
like DieHard, we exhibit much better variance and overall performance.

The second series of benchmarks stress-test the allocator via specific allocation patterns which are not
representative of real-world applications. We remark that for most of these,
\name performs within 30\% of \hm.
One notable exception is rptestN, which particularly tests large allocations, and exhibits the same
limitations as \li+redis+.

Since we inherit most of the design choices from \hm, our performance is further
apart from a non-security-oriented allocator like glibc. This is the
cost of defensive security measures; ultimately, application developers will
have to decide how much importance they attach to the security of their applications.
Nevertheless, \name outperforms mng for some benchmarks.

Broadening our scope to look at other allocators (\tref{perf-full}), we observe
a greater variety of performance profile across allocators, with greater variation
on micro-benchmarks than on realistic workloads.
These results suggest that
no single choice of optimizations is the ``right'' one for every
workload. Additional benchmarks (not listed here) on different processors show that
microarchitecture revisions can also significantly influence the final results.
We believe that by addressing the long tail of micro-optimizations (such as carefully annotating each
branch with \li+likely+/\li+unlikely+ attributes, optimizing size class selection, or supporting
larger slab allocations), we can meet the performance of \hm. In short, on most benchmarks, our
verified code is almost as efficient as the unverified one, and for a few corner cases, the
performance can be regained with many time-consuming small fixes which we intend to enact.

\myparagraph{Memory usage}
We measure the Resident Set Size (RSS), with swap disabled to get an accurate
reading of the memory footprint of a running program. RSS only counts pages that
are actually used, as opposed to the \emph{virtual} memory usage; this is what
we want.
We observe a wide variety of behaviors; in several cases, \name outperforms
\hm, while in others, its memory footprint is larger than \hm.
While our allocator design is inspired by \hm, we posit that this variance is
due to minor implementation differences, especially in the quarantine.
We remark that glibc's allocator, too, exhibits suboptimal behavior compared to \hm
on several test cases, despite the absence of a quarantine mechanism,
which greatly increase the memory usage;
this again suggests that no design is ideal for all applications.

\subsection{Security Evaluation}
We adopt the state-of-the-art security-oriented allocator design from \hm, and as such, inherit
  its security guarantees. Out of due diligence, we confirm that our implementation, unsurprisingly,
  is capable of defending against a wide variety of attacks, and as such, provides an efficient
``last line of defense'' against exploitation and programmer mistakes.
We tested various use-after-frees (UAFs) found in recent fuzzing-based
work~\cite{yagemannpumm}, and confirmed that they were indeed caught. We tested
a ``classic'' CVE~\cite{cvebzip2} for a UAF in bzip2, also prevented by \name.
We tested a series of exploits from the how2heap
repository~\cite{how2heap}. We prevent the ``fastbin\_reverse\_into\_cache'', ``fastbin\_dup'' and
``tcache poisoning'' exploits, all of which are successful with the most recent
version of glibc (2.38). We thus conclude that the security measures we
implemented are, as we expected, effective.
We re-emphasize that use-after-frees and double-frees are still very much active attack vectors,
and continue to pop up in mainstream software~\cite{cve-openssh-2023,cve-curl-2023}.

\subsection{Firefox Integration}
To assess whether \name can be deployed as a drop-in replacement into real-world projects,
we integrated it into Mozilla Firefox, raising interesting challenges: first, the replacement allocator
must to provide non-standard extensions of \ls`malloc` and \ls`free`~\cite{glandium2012};
second Firefox, and hence the underlying allocator, must work in a highly concurrent setting,
supporting many threads and processes in parallel.
To evaluate the user-facing impact of using \name, we relied on \citet{JetStream2}, a ``benchmark suite focused on the most advanced web applications''.
We integrated \name with Firefox 118.0.2 and successfully ran the entire benchmark suite, demonstrating a
performance overhead of 5.1\% and 12.1\% compared to a version compiled with \hm and glibc respectively; this
suggests the applicability of \name to real-world usecases.

\subsection{Proof Effort}
\label{sec:proofeffort}

\begin{wraptable}{r}{0.3\textwidth}
  \vspace{-0.6cm}
  \centering
  \small
  \begin{tabular}{l|r}
    Component & LoC \\
    \hline
    \textbf{Reusable Libraries} \\
    \quad Bitmaps & 1,520 \\
    \quad List & 2,754 \\
    \quad Queue & 494 \\
    \quad AVL & 4,283 \\
    \quad Misc & 3,692 \\
    \textbf{\name} & 12,972 \\
    \textbf{C glue} & 303 \\
    \textbf{Total} & 26,102
  \end{tabular}
  \vspace{0.2cm}
  \small
  \caption{Lines of \fstar code for each component of \name, excluding whitespace and
  comments. Misc includes
  \li+starseq+, other helpers related to \li+sldep+ and \li+slrefine+, plus
  various library functions not found in \fstar's standard library.}
  \label{tab:loc}
  \vspace{-0.5cm}
\end{wraptable}

We estimate that our implementation effort for \name took two person-years. This
includes building a substantial foundation that was missing from the Steel
project, starting from arrays, and moving up to data structures (AVL trees,
bitmaps, arraylists, queues). A lot of work was foundational, and aimed at 
understanding how to successfully model, verify and extract an extremely
low-level view of memory within the framework. We believe that this effort will
be highly reusable, and intend to upstream it as soon as possible.

We reiterate that a substantial amount of effort went into building strong,
tight abstraction boundaries between the various components of \name.
In less than a week, we implemented a first batch of security mechanisms without affecting the rest of the code. More recently, we improved the quarantine mechanism, relying on the tight abstraction boundaries to preserve verification and extraction.
The tight abstractions have the added benefit of preserving a lot of intermediary
verification files, meaning that one can efficient rebuild and re-extract \name
when iterating on one sub-component.

\section{Future Work and Conclusion}

\name is a verified, efficient, security-oriented memory allocator that implements
state-of-the-art performance and defensive design choices. Using the \fstar
proof assistant and the Steel separation logic framework,
we verified, implemented and extracted to C a complete allocator
that can be used as a drop-in replacement for realistic workloads. To the best
of our knowledge, this is the first concurrent allocator that is simultaneously verified,
hardened, and efficient.

Most of our future work centers on closing the gap between \name and \hm (our
chief source of inspiration), both in terms of performance and breadth of
security features.
Overall, we believe that there is currently a gap in real-world verified
software stacks when it comes to allocators, and we hope that
\name can represent a first step towards filling that gap.

\begin{acks}
We would like to thank Daan Leiijen for clarifications about many parts of mimalloc and mimalloc-bench, Matthew Parkinson for discussions about allocators security mechanisms, and Tim Bourke for valuable feedback about earlier versions of this paper.
\end{acks}

\bibliographystyle{ACM-Reference-Format}
\bibliography{conferences,cited}

\appendix

\section{Performance Benchmarks}

\begin{table*}
  \centering
  \small
  \begin{tabular}{l|r|r|r|r|r|r|r|r|r|r|r}
Benchmark & dh & ff & hm & iso & mi-sec & mng & scudo & sg & sn-sec & st & glibc \\ \hline
barnes & 1.036 & 1.005 & 1.0 & 1.003 & 1.003 & 1.0 & 1.005 & 1.0 & 1.0 & 1.0 & 1.0\\
cfrac & 1.446 & 0.762 & 1.0 & 0.986 & 0.646 & 0.738 & 0.675 & 0.541 & 0.528 & 1.278 & 0.544\\
espresso & 1.273 & 0.917 & 1.0 & 0.999 & 0.736 & 1.301 & 0.753 & 0.707 & 0.642 & 2.284 & 0.707\\
gs & 2.278 & 1.157 & 1.0 & 0.435 & 0.357 & 0.678 & 0.374 & 0.339 & 0.452 & 1.304 & 0.339\\
larsonN & 17.097 & 2.164 & 1.0 & 6.254 & 0.144 & 2.251 & 0.215 & 0.09 & 0.073 & 1.036 & 0.09\\
larsonN-sized & 16.823 & 2.121 & 1.0 & 6.165 & 0.141 & 2.2 & 0.211 & 0.088 & 0.072 & 1.008 & 0.088\\
leanN & 6.249 & 0.707 & 1.0 & 6.81 & 0.602 & 1.421 & 0.666 & 0.615 & 0.515 & 1.086 & 0.617\\
lua & TO & 1.382 & 1.0 & 0.883 & 0.768 & 1.05 & 0.755 & 0.728 & 0.724 & 1.554 & 0.729\\
mathlib & ERR & 0.726 & 1.0 & 1.258 & 0.652 & 0.779 & 0.686 & 0.649 & 0.562 & 1.221 & 0.645\\
redis & 2.809 & 0.764 & 1.0 & 1.449 & 0.919 & 0.82 & 0.638 & 0.59 & 0.559 & 16.379 & 0.592\\
rocksdb & ERR & 0.921 & 1.0 & ERR & 0.903 & ERR & 0.874 & 0.829 & ERR & 0.843 & 0.813\\
z3 & 5.063 & 0.813 & 1.0 & 1.0 & 0.75 & 0.813 & 0.75 & 0.688 & 0.625 & 1.375 & 0.688\\
\hline
alloc-test1 & 2.051 & 0.757 & 1.0 & 1.096 & 0.587 & 0.922 & 0.569 & 0.521 & 0.426 & 1.291 & 0.515\\
alloc-testN & 12.566 & 0.288 & 1.0 & 5.185 & 0.158 & 3.693 & 0.166 & 0.135 & 0.112 & 1.574 & 0.136\\
cache-scratch1 & 1.006 & 1.006 & 1.0 & 1.0 & 0.994 & 0.994 & 0.994 & 0.994 & 0.994 & 0.994 & 0.994\\
cache-scratchN & 0.977 & 1.023 & 1.0 & 1.0 & 0.977 & 0.977 & 0.977 & 0.977 & 0.977 & 1.0 & 0.977\\
cache-thrash1 & 1.006 & 1.05 & 1.0 & 1.0 & 0.994 & 0.994 & 0.994 & 0.994 & 0.994 & 1.0 & 0.994\\
cache-thrashN & 1.0 & 1.023 & 1.0 & 1.186 & 1.0 & 1.0 & 1.0 & 1.0 & 1.0 & 1.0 & 1.0\\
glibc-simple & 1.5 & 0.632 & 1.0 & 0.976 & 0.519 & 1.187 & 0.567 & 0.456 & 0.313 & 1.931 & 0.453\\
glibc-thread & 10.033 & 0.284 & 1.0 & 5.295 & 0.11 & 3.161 & 0.12 & 0.071 & 0.061 & 1.679 & 0.074\\
malloc-large & 3.604 & 1.036 & 1.0 & 0.58 & 0.538 & 1.034 & 1.036 & 0.539 & 0.545 & 1.046 & 0.539\\
mleak10 & 1.692 & 5.0 & 1.0 & 1.0 & 0.923 & 0.923 & 1.0 & 1.231 & 0.923 & 1.0 & 1.231\\
mleak100 & 1.754 & 5.566 & 1.0 & 1.041 & 1.016 & 1.0 & 1.057 & 1.369 & 0.975 & 1.041 & 1.369\\
mstressN & 5.251 & 0.812 & 1.0 & 2.677 & 0.462 & 1.722 & 0.7 & 0.413 & 0.354 & 1.605 & 0.435\\
rbstress1 & 1.14 & 0.85 & 1.0 & 0.984 & 0.863 & 0.809 & 0.764 & 0.817 & 0.779 & 0.895 & 0.812\\
rbstressN & 1.285 & 0.757 & 1.0 & 1.051 & 0.797 & 0.782 & 0.755 & 0.805 & 0.883 & 0.939 & 0.76\\
rptestN & 7.521 & 3.404 & 1.0 & 2.432 & 0.471 & 1.414 & 1.015 & 0.397 & 0.535 & 19.537 & 0.387\\
sh6benchN & 22.109 & 0.305 & 1.0 & 7.598 & 0.129 & 3.239 & 0.502 & 0.173 & 0.074 & 1.649 & 0.175\\
sh8benchN & 13.098 & 0.302 & 1.0 & 9.578 & 0.121 & 2.229 & 0.648 & 0.134 & 0.063 & 1.256 & 0.133\\
xmalloc-testN & 8.978 & 0.298 & 1.0 & 3.559 & 0.116 & 1.543 & 1.887 & 0.204 & 0.062 & 1.17 & 0.205\\
\end{tabular}

  \caption{
  Execution time of various programs, measured against the
  following allocators:
  dieharder (``dh'', \cite{novark2010dieharder}, revision 640949f),
  FFmalloc (``ff'', \cite{ffmalloc}, revision 2f24ecf),
  \hm default version (``hm'', \cite{hardenedmalloc}, revision 995ce07),
  isoalloc (``iso'', \cite{isoalloc}, revision 2670d5f),
  mimalloc-secure (``mi-sec'', \cite{leijen2019mimalloc}, revision b66e321),
  nextgen malloc implementation for musl libc (``mng'', \cite{muslmalloc}, revision 2ed5881),
  scudo (``scudo'', \cite{scudoallocator}, revision 1654d7d),
  slimguard (``sg'', \cite{slimguard}, revision 7d9139a),
  snmalloc-secure (``sn-sec'', \cite{snmalloc}, revision dc12688),
  \name (``st'', this work),
  and the glibc allocator (``glibc'', version 2.38).
  All times are normalized and presented relative to \hm. Benchmarks ending in N
  exercise very narrow patterns of allocation which stress-test the allocator
  and are not representative of a general usage pattern. \name exhibits pathologically bad
  performance on the rptestN and redis benchmarks, which particularly test large allocations,
  currently under-optimized in \name. ERR corresponds to crashes during benchmarking, and TO represents cases that exceeded the time allocated for benchmarking}
  \label{tab:perf-full}
\end{table*}

\begin{table*}
  \centering
  \small
  \begin{tabular}{l|r|r|r|r|r|r|r|r|r|r|r}
Benchmark & dh & ff & hm & iso & mi-sec & mng & scudo & sg & sn-sec & st & glibc \\ \hline
barnes & 0.774 & 1.158 & 1.0 & 0.959 & 0.753 & 0.745 & 0.702 & 0.686 & 0.74 & 0.832 & 0.699\\
cfrac & 0.261 & 8.595 & 1.0 & 1.046 & 0.15 & 0.1 & 0.181 & 0.111 & 0.124 & 0.81 & 0.113\\
espresso & 0.18 & 0.924 & 1.0 & 1.203 & 0.144 & 0.066 & 0.122 & 0.063 & 0.327 & 1.399 & 0.062\\
gs & 0.797 & 1.876 & 1.0 & 1.383 & 0.766 & 0.524 & 0.553 & 0.517 & 0.743 & 1.39 & 0.513\\
larsonN & 0.659 & 2.213 & 1.0 & 0.921 & 0.906 & 0.369 & 0.462 & 0.69 & 1.137 & 2.277 & 0.664\\
larsonN-sized & 0.69 & 2.251 & 1.0 & 0.933 & 0.938 & 0.351 & 0.469 & 0.676 & 1.129 & 2.292 & 0.78\\
leanN & 1.421 & 7.883 & 1.0 & 2.834 & 1.117 & 0.688 & 0.856 & 0.754 & 1.153 & 1.417 & 0.78\\
lua & TO & 1.077 & 1.0 & 1.334 & 0.764 & 0.639 & 0.753 & 0.645 & 0.794 & 1.338 & 0.654\\
mathlib & ERR & 2.25 & 1.0 & 4.711 & 1.071 & 0.626 & 0.718 & 0.66 & 0.891 & 2.004 & 0.676\\
redis & 0.493 & 2.095 & 1.0 & 1.166 & 0.369 & 0.216 & 0.255 & 0.207 & 0.301 & 1.339 & 0.205\\
rocksdb & ERR & 1.859 & 1.0 & ERR & 1.03 & ERR & 0.755 & 0.774 & ERR & 1.471 & 0.785\\
z3 & 0.622 & 1.593 & 1.0 & 1.511 & 0.615 & 0.417 & 0.459 & 0.4 & 0.646 & 1.294 & 0.401\\
\hline
alloc-test1 & 0.447 & 16.602 & 1.0 & 2.122 & 0.42 & 0.292 & 0.362 & 0.305 & 0.368 & 1.471 & 0.305\\
alloc-testN & 0.302 & 19.276 & 1.0 & 0.92 & 0.611 & 0.247 & 0.285 & 0.201 & 0.295 & 2.678 & 0.201\\
cache-scratch1 & 0.205 & 2.737 & 1.0 & 1.07 & 0.288 & 0.133 & 0.141 & 0.133 & 0.141 & 0.445 & 0.133\\
cache-scratchN & 0.22 & 2.933 & 1.0 & 1.227 & 0.309 & 0.144 & 0.152 & 0.146 & 0.153 & 0.555 & 0.159\\
cache-thrash1 & 0.219 & 2.94 & 1.0 & 1.151 & 0.308 & 0.143 & 0.149 & 0.142 & 0.15 & 0.401 & 0.143\\
cache-thrashN & 0.204 & 2.735 & 1.0 & 1.072 & 0.285 & 0.135 & 0.141 & 0.135 & 0.141 & 0.517 & 0.133\\
glibc-simple & 0.248 & 1.343 & 1.0 & 1.119 & 0.201 & 0.055 & 0.206 & 0.058 & 0.086 & 0.744 & 0.058\\
glibc-thread & 0.133 & 0.547 & 1.0 & 0.599 & 0.458 & 0.032 & 0.096 & 0.057 & 0.428 & 2.813 & 0.056\\
malloc-large & 0.928 & 1.084 & 1.0 & 4.295 & 1.288 & 0.918 & 0.918 & 1.197 & 1.547 & 0.928 & 1.196\\
mleak10 & 0.284 & 2.683 & 1.0 & 1.018 & 0.294 & 0.07 & 0.156 & 0.143 & 0.1 & 0.682 & 0.146\\
mleak100 & 0.258 & 12.671 & 1.0 & 0.974 & 0.266 & 0.062 & 0.141 & 0.139 & 0.09 & 0.618 & 0.128\\
mstressN & 0.811 & 1.041 & 1.0 & 2.407 & 1.774 & 0.681 & 0.755 & 1.561 & 1.329 & 1.12 & 1.521\\
rbstress1 & 1.177 & 1.093 & 1.0 & 2.0 & 0.868 & 0.905 & 1.004 & 0.708 & 0.802 & 0.925 & 0.708\\
rbstressN & 0.705 & 1.023 & 1.0 & 1.803 & 1.277 & 0.684 & 0.76 & 1.096 & 1.062 & 0.941 & 1.0\\
rptestN & 0.534 & 0.736 & 1.0 & 0.81 & 0.611 & 0.234 & 0.257 & 0.265 & 0.699 & 2.59 & 0.254\\
sh6benchN & 1.771 & 5.287 & 1.0 & 5.416 & 0.901 & 0.868 & 1.189 & 1.061 & 0.987 & 1.12 & 1.056\\
sh8benchN & 1.075 & 1.004 & 1.0 & 2.673 & 1.39 & 0.803 & 0.774 & 0.996 & 1.259 & 1.21 & 0.982\\
xmalloc-testN & 0.836 & 0.956 & 1.0 & 0.536 & 0.625 & 0.066 & 0.746 & 0.806 & 0.564 & 0.393 & 0.815\\
\end{tabular}

  \caption{Resident Set Size (RSS) memory usage for various programs,
  measured against the same set of allocators as \tref{perf-full}. Numbers are, as
  before, normalized and presented relative to \hm.}
\end{table*}

\end{document}